\documentclass[amsmath,amssymb,floatfix,twocolumn]{revtex4-1}
\usepackage{graphicx}
\usepackage{rotating}
\usepackage{longtable}
\usepackage{supertabular}
\usepackage{latexsym}
\usepackage{amssymb}
\usepackage{float}
\usepackage{dcolumn}
\usepackage{bm}
\usepackage{amsmath}

\begin{document}

\title{Performance of Wang-Landau algorithm in lattice model of liquid crystals}
\author{Suman Sinha}
\email{suman.sinha.phys@gmail.com, Tel : 91 9433 260512, Fax : 91 33 2351 9755}
\affiliation {Department of Physics, University of Calcutta,\\
92 Acharya Prafulla Chandra Road, \\
Kolkata - 700009, India}

\begin{abstract}
We present a study on the performance of Wang-Landau algorithm in a lattice model of 
liquid crystals which is a continuous lattice spin model. We propose a novel method
of the spin update scheme in a continuous lattice spin model. The proposed scheme 
reduces the autocorrelation time of the simulation and results in faster convergence.
\\

\end{abstract}


\maketitle

\section{Introduction}
\label{intro}
The Wang-Landau (WL) algorithm \cite{wl}, introduced in 2001, has received much attention and has been applied
to a wide range of problems \cite{yo,oty,sdp,rp,jp,yfp,jsm,cp,lwll,wl1,swl}. In most of these investigations, the
authors have applied the WL algorithm to systems with discrete energy levels. However, relatively fewer papers
have so far appeared on lattice models with continuous energy spectrum \cite{yfp,pcabd,zstl,mgr,br,sr}. Techniques,
 in general, to improve the algorithm for different problems have also been proposed \cite{yk,sbm,sbml,bab,yp,sdp2,
tht,dtwwtsc,vmmb,zb,td,lol,ed,ml,bp,zs}. The review \cite{lte} illustrates the versatile applications of the WL algorithm
in protein folding, fluid simulations, systems with first order phase transitions and other systems with rough 
energy terrain. Some authors find its applications in performing numerical integration \cite{td,lwll}.

The WL algorithm allows us to calculate the density of states (DOS) as a function of energy or the joint density of
states (JDOS) as a function of energy and a second variable \cite{zstl}. For a macroscopic system, the DOS
$\Omega(E_i)$ (where $i=1,2,\cdots ,n$, $n$ being the bin index) is a large number and it is convenient to work 
with its logarithm $g(E_i)={\tt \ln} ~\Omega(E_i)$. Since the DOS is independent of temperature and contains
complete information about the system, the task is to determine it as accurately as possible. The next step
involves the determination of partition function \smash{$Z(T)= \sum_E \Omega(E)~ {\tt e}^{-\beta E}$} 
($\beta = 1/T$, Boltzmann constant has been set to unity) at any temperature ($T$) by the standard 
Boltzmann reweighting procedure. Once the partition function is known, the model is essentially ``solved'' 
since most thermodynamic quantities at any temperature can be calculated from it. The algorithm is implemented by performing an 
one-dimensional random walk that produces a ``flat'' histogram in the energy space. For a continuous model, one needs to
use a discretization scheme to divide the energy range of interest into a number of bins which label the 
macrostates of the system. In the WL algorithm, these macrostates are sampled with a probability which is
proportional to the reciprocal of the current DOS. The estimate for the DOS is improved at each step of the
random walk using a carefully controlled modification factor $f$ to produce a result that converges to the
true DOS quickly. A histogram record $H(E_i)$ of all states visited is maintained throughout the 
simulation. When $g(E_i)$ corresponding to a certain macrostate is modified as $g(E_i) \rightarrow
g(E_i)+{\tt \ln}f$, the corresponding $H(E_i)$ is modified as $H(E_i) \rightarrow H(E_i)+1$. In the 
original proposal of WL algorithm, an iteration is said to be complete when the histogram satisfies a 
certain ``flatness'' condition. This means that $H(E_i)$, for all values of $i$, has attained $90\%$
(or some other preset value) of the average histogram. In the following iteration, $f$ is reduced in
some fashion, the $H(E_i)$'s are reset to zero and the process is continued till ${\tt \ln}f$ is as
small as $10^{-8}$ or $10^{-9}$. Since the history of the entire sampling process determines the DOS, the
WL algorithm is non-Markovian besides being multicanonical in nature.

In course of the random walk in a WL simulation, the fluctuations of energy histogram, for a given
modification factor $f$, initially grows with time and then saturates to a certain value. Zhou and
Bhatt \cite{zb} carried out a mathematical analysis of the WL algorithm. They provided a proof of the convergence
of the iterative procedure and have shown that the fluctuations in histogram, proportional to 
$1/ \sqrt {\ln f}$ for a given $f$, cause statistical errors which can be reduced by averaging over 
multiple simulations. They have also shown that the correlation between adjacent records in the 
histogram introduces a systematic error which is reduced at smaller $f$. The prediction in Ref.
\cite{zb} has been numerically verified by different authors independently \cite{lol,sr}. Although 
to obtain a flat histogram is the initial motivation behind the WL algorithm, Ref. \cite{zb} concluded
that flatness is not a necessary criterion to achieve convergence and suggested that one should
instead focus on the fluctuations of the histogram rather than the ``flatness''. 
They had shown that $1/ \sqrt {\ln f}$ visits on each macroscopic state is enough to guarantee the
convergence. In fact, fluctuations 
in the histogram is intrinsic to WL algorithm. These fluctuations lead to a statistical error in the 
DOS which scales as $\sqrt {\ln f}$, for a given $f$. The iterative WL algorithm partially reduces this
statistical fluctuations by decreasing $f$ monotonically. However Ref. \cite{bp} clearly 
illustrates that even if $f$ is reduced to a very small value according to the original prescription, the
statistical error stops to decrease at a certain point. In practice there always exists a systematic 
error in the simulation which is a function of $f$ and the correlation between adjacent records in the
histogram. Ref. \cite{zb} observed that this systematic error decreases when either $f$ or the 
correlation decreases. In this context, we refer to the work of Morozov and Lin \cite{ml} who presented
a study on the estimations of accuracy and convergence of the Wang-Landau algorithm on a two level system
with a significant efficiency improvement in \cite{ml2}.
The WL algorithm compares $\Omega (E_i)$ and $\Omega (E_f)$, i.e, DOS before and after an
attempted move, but it does not require $E_i$ to be close to $E_f$. This is why Ref. \cite{zb} suggested the
use of cluster algorithms that allow ``nonlocal'' moves in the parameter space. The Ref. \cite{zs} 
rightly pointed out that the update schemes for the underlying model certainly have an effect on the
outcome. In the present paper we suggest a method for the spin update scheme of a lattice model with 
continuous energy spectrum, which reduces the autocorrelation time by an appreciable amount compared to the
conventional spin update scheme. The suggested spin update method to obtain a less correlated 
configuration has also the advantage that this method is free from tuning any adjustable parameter. The
method is described in Section \ref{ct}. We also investigate the growth of the histogram fluctuations
in the one-dimensional Lebwohl-Lasher (LL) model, described in Section \ref{model},
 to check if the nature of the dependence of the maximum of the histogram 
fluctuations on the modification factor $f$ is model independent or not. We mention in passing that 
Ref. \cite{lol} suggested the model-independent nature of the maximum of the histogram fluctuations 
by performing simulations on two discrete Ising models and concluded that many more simulations on
different models are needed to confirm this universality nature. Ref. \cite{sr} confirmed this 
universality behavior for two continuous lattice spin models with spin dimensionality two.
We have found that for the present model (spin dimensionality three),
the fluctuations in the energy histogram, after an initial increase, saturates to a value which
is inversely proportional to $\sqrt {\ln f}$ and confirm that this feature is generic to the WL algorithm.
In the second part of the work, we have carried out the WL simulation with the proposed spin update scheme
to estimate the canonical averages of various thermodynamic
quantities for lattices of reasonably large size where minimum number of visits to each macrostate are
$1/ \sqrt{\ln f}$. Results obtained from our simulation are compared with the exact results available 
for the model.

The rest of the paper is arranged as follows. In Section \ref{model}, we have described the model. The
computational techniques are discussed in Section \ref{ct}. Section \ref{rd} presents our results and discussions. 
 Section \ref{conclu} draws the conclusions.

\section{Model}
\label{model}
For the purpose of investigation, we have chosen an one-dimensional array of three-dimensional spins
($d=1,l=3$, where $d$ is the space dimensionality and $l$ is the spin dimensionality) interacting 
with nearest neighbors (nn) via a potential
\begin{equation}
V_{ij}=-P_2(\tt \cos~ \theta_{ij})
\label{eqn1}
\end{equation}
where $P_2$ is the second Legendre polynomial and $\theta_{ij}$ is the angle between the nearest
neighbor spins $i$ and $j$ (the coupling constant in the interaction has been set to unity). The
spins are three-dimensional and headless, i.e, the system has the $O(3)$ as well as the local $Z_2$
symmetry, characteristic of a nematic liquid crystal. The model, known as the Lebwohl-Lasher (LL) 
model \cite{ll}, is the lattice version of the Maier-Saupe (MS) model \cite{ms} which describes a 
nematic liquid crystal in the mean field approximation. Being a low-dimensional model with nn
interaction, the $1d$ LL model does not exhibit any finite temperature phase transition. This model
has been solved exactly by Vuillermot and Romerio \cite{vr} in $1973$, using a group theoretical
method. The results obtained in \cite{vr} are quoted below.
The partition function $Z_N(\widetilde K)$ for the $N$-particle system is given by
\begin{equation}
Z_N(\widetilde K)=\widetilde K^{N/2} {\tt exp}\left[\frac {2}{3}N\widetilde K \right]D^N(\widetilde K^{1/2})
\label{eqn2}
\end{equation}
where $\widetilde K=3/2T$ is a dimensionless quantity. $D$ is the Dawson function \cite{as} given by
\begin{equation}
D(x)={\tt exp}(-x^2)\int_0^x e^{u^2}du
\label{eqn3}
\end{equation}
The dimensionless internal energy $U_N(\widetilde K)$, entropy $S_N(\widetilde K)$ and the specific heat
$C_N(\widetilde K)$ are given by
\begin{equation}
\frac{2U_N(\widetilde K)}{N}=1+\frac{3\widetilde K^{-1}}{2}-\frac{3}{2} \widetilde K^{-1/2}D^{-1}
\left (\widetilde K^{1/2}\right)
\label{eqn4}
\end{equation}
\begin{multline}
\frac{S_N(\widetilde K)}{N}=\frac{1}{2}+\widetilde K-\frac{1}{2}\widetilde K^{1/2}D^{-1}
\left (\widetilde K^{1/2}\right) \\
+\ln \left[\widetilde K^{-1/2}D\left(\widetilde K^{1/2}\right)\right]
\label{eqn5}
\end{multline}
\begin{multline}
\frac{2C_N(\widetilde K)}{N}=1-\widetilde K^{3/2}\left[\frac{\widetilde K^{-1}}{2}-1\right]D^{-1}
\left (\widetilde K^{1/2}\right) \\
-\frac{1}{2} \widetilde K D^{-2}\left(\widetilde K^{1/2}\right)
\label{eqn6}
\end{multline}
We decided to choose this model to test the performance of WL algorithm using the suggested spin update
scheme so that a comparison can be made with the exact results available for the model.

\section{Computational Techniques}
\label{ct}
In the first part of this Section, we will describe the computational techniques used to determine the fluctuations
in the energy histogram. In the later part of this Section, we will discuss the method for the new spin update scheme.

Let us first explain the notations and symbols relevant to the present work. The saturation value of the 
energy histogram fluctuation in the $k^{th}$ iteration is represented by $\beta_k$. Let $f_k$ be the 
modification factor for the $k^{th}$ iteration. One usually starts with a modification factor $f=f_1 \geq 1$ 
and uses a sequence of decreasing $f_k$'s ($k=1,2,3,\cdots$) defined in some manner. One Monte Carlo (MC) sweep is taken 
to be completed when the number of attempted single spin moves equals the number of spins in the system. The error 
in the DOS after the $n^{th}$ iteration is directly related to $\beta_i$ for $i > n$, the saturation values of
the fluctuations.
In the WL algorithm the logarithm of the DOS after $n$ iterations is given by
\begin{equation}
g_n(E_i)=\sum_{k=1}^n H_k(E_i)\ln (f_k)
\label{eqn7}
\end{equation}
where $H_k(E_i)$ is the accumulated histogram count for the $i^{th}$ energy bin during the $k^{th}$ iteration.
In order to get an idea of the fluctuations in the histogram and its growth with the number of MC sweeps, we subtract
the minimum of the histogram count $h_k^j$ which occurs in the histogram after the $j^{th}$ MC sweep has been
completed during the $k^{th}$ iteration, i.e., we consider the quantity
\begin{equation}
\widetilde H_k^j(E_i)=H_k^j(E_i)-h_k^j
\label{eqn8}
\end{equation}
It may be noted that $h_k^j$ does not refer to any particular bin and may occur in any of the
visited bins. The quantity $\widetilde H_k^j(E_i)$ is now summed over all bins to give $\Delta H_k^j$.
\begin{equation}
\Delta H_k^j=\sum_i \widetilde H_k^j(E_i)
\label{eqn9}
\end{equation}
$\Delta H_k^j$ is thus a measure of the fluctuations which occurs in the $j^{th}$ MC sweep during
$k^{th}$ iteration and is a sort of average over all macrostates or bins. $\Delta H_k^j$ fluctuates 
with $j$ because of statistical errors and its mean value taken over $j$ is nothing but $\beta_k$. The
error of the logarithm of the DOS, summed over all energy levels or bins, after the completion of $n$
iterations is therefore given by \cite{lol}
\begin{equation}
\eta_n=\sum_{k=n+1}^{\infty} \beta_k \ln (f_k)
\label{eqn10}
\end{equation}
Eq. (\ref{eqn10}) means that the error depends only on the fluctuations in histogram and the 
sequence of modification factors. When the values of $f_k$ are predetermined, the fluctuations 
in histogram, i.e., $\Delta H_k^j$, becomes the only determining factor for the error. For this
reason the observable $\Delta H_k^j$, defined by Eq. (\ref{eqn9}), is considered to be a good measure of
the fluctutations in histogram. However, we point out that because of the summation over the index $i$
in Eq. (\ref{eqn9}), the nature of the distribution of the errors over the energy bins is not 
reflected in the summed quantity $\Delta H_k^j$. What we get instead is an error which has been summed 
over all the energy bins.
Since the predicted value of the error $\eta_n$ is of the order of $\sqrt {\ln f_n}$ \cite{zb}, one
expects that the histogram saturation value $\beta_n$, for the $n^{th}$ iteration, should be 
proportional to $1/ \sqrt {\ln f_n}$.
\subsection{Proposal for a novel spin update method}
Now we discuss the method to generate a subsequent less-correlated spin configuration. In the 
conventional spin update method for a continuous lattice spin model, the orientation of each
spin $\vec s$ is stored in terms of the direction cosines $(l_1, l_2, l_3)$. To generate a 
new configuration (microstate), a spin is selected at random and each direction cosine of it
is updated as $l_i \rightarrow l_i +p*x_i$ for ($i=1,2,3$) where the parameter ``p'' denotes
the amplitude of the random angular displacements, chosen such that approximately half of the 
configurations are accepted and half rejected \cite{lg} and $x_i$ is a random number 
between $-1$ to $1$. We have seen for a number of continuous lattice spin models that the
results for the thermodynamic quantities become very sensitive to the value of the parameter
``p''. ``p'' is generally taken such that $p < 1$ and the choice of ``p'' also depends on the
systems we are working on. The reason for taking $p < 1$ is that small values of ``p'' 
correspond to small changes in the direction of the spin, i.e., the energy cost of an attempted
move will be small. However, this is not the only form of update, nor is it known whether this
is the most efficient form. The thing is, there is quite a lot of flexibility about the choice 
of the new state for the spins. A good discussion of it may be found in Ref. \cite{nb}.

In the present work, we propose a novel protocol to generate a less-correlated spin configuration
in the following manner. We take a random unit vector $\vec r$ and a spin update $\vec s \rightarrow \vec s^{~\prime}$ 
is defined as $\vec s^{~\prime}=\vec s - 2\left(\vec s \cdot \vec r\right)\vec r$ where $(\vec s \cdot \vec r)$ 
is the dot product of $\vec s$ and $\vec r$. This represents a reflection with respect to the hyperplane 
orthogonal to $\vec r$ and this is an idempotent operation. The idea came from Wolff \cite{wlf}. One 
may think of a linear transformation $R(\vec r)$ such that $\vec s^{~\prime}=R(\vec r) \vec s$. This 
linear transformation has the property 
\begin{equation}
R(\vec r)^2=1
\end{equation}
i.e., idempotent and
\begin{equation}
[R(\vec r)\vec s_1]\cdot[R(\vec r)\vec s_2]=\vec s_1 \cdot \vec s_2
\end{equation}
i.e., the Hamiltonian (\ref{eqn1}) is invariant under global R transformations. 
 This spin update method reduces the autocorrelation time to a considerable amount and consequently 
systematic error decreases. Moreover, 
defining a spin update in that way, the algorithm becomes free from tuning any adjustable parameter
even while simulating a lattice spin model with continuous energy spectrum. This spin update method has
resulted in efficient simulation of continuous lattice spin models with $XY$ symmetry \cite{sr3,sr4}.

The energy of the $1d$ LL model is a continuous variable and it can have any value between $-L$ to 
$L/2$ where $L$ is the system size. To discretize the system, we have chosen an energy range 
($-L,0$) and divided this energy range into a number of bins (macrostates) each having a width, say
$w$. In the present work, the bin width is taken to be $0.2$.

\section{Results and discussions}
\label{rd}
\begin{figure*}
\begin{center}
\begin{tabular}{cc}
      \resizebox{85mm}{!}{\includegraphics[scale=0.6]{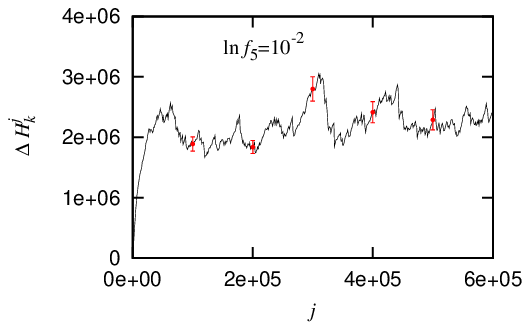}} &
      \resizebox{85mm}{!}{\includegraphics[scale=0.6]{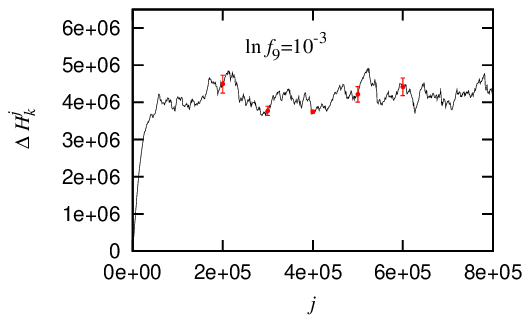}} \\
      \resizebox{85mm}{!}{\includegraphics[scale=0.6]{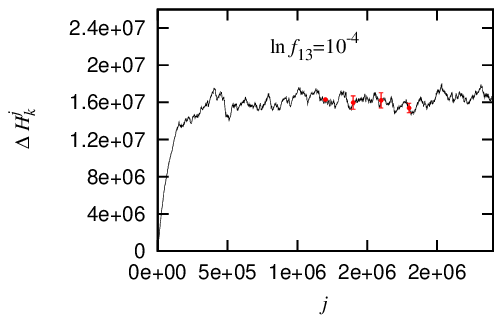}} &
      \resizebox{85mm}{!}{\includegraphics[scale=0.6]{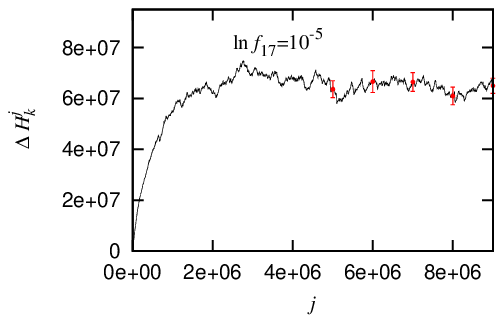}} \\
    \end{tabular}
\end{center}
\caption{(Color online) The histogram fluctuations $\Delta H_k^j$ for the $k^{th}$ iteration are plotted
against the MC sweeps $j$ for the $L=160$ lattice. The values of $\ln f_k$ are indicated
in the figures. The histograms are averaged over $100$ independent simulations.}
\label{hisfluc}
\end{figure*}
We have determined for the lattice model we have defined, the dependence of the quantity
$\Delta H_k^j$, given by Eq. (\ref{eqn9}), on $j$, the number of MC sweeps for a given
iteration denoted by $k$. For the purpose of testing the fluctuations in histogram, we 
have taken linear spin chains of length $L=80$ and $160$. Nearest neighbor interactions 
along with periodic boundary conditions were always used. The starting value of the 
modification factor $\ln f_1$ was taken to be $0.1$ and the sequence $\ln f_{n+1}=
(\ln f_n)/10^{1/4}$ was chosen and for the purpose of determination of fluctuations, 
the minimum $\ln f$ used was $10^{-5}$. Clearly, the chosen sequence of $f$ is to ensure
that it gets reduced by a factor of $10$ after four iterations. We have determined the
quantity $\Delta H_k^j$ defined by Eq. (\ref{eqn9}) at intervals of $10^3$ MC sweeps and
the maximum number of sweeps chosen for a given value of $f$ is such that the saturation
of the histogram is clearly evident. The system energy is always considered up to 
$E=0$. The lower limit of the energy for $L=80$ is taken to be $-78$ and for $L=160$, it
is $-158$, while the corresponding ground state energies are $-80$ and $-160$. Thus the
visited energy range goes to a sufficiently low value to cover the entire range of interest, 
though the small cut near the ground state is necessary, as configurations near the 
minimum energy take a very long time to be sampled during the random walk.

\begin{figure}[!h]
\begin{center}
\resizebox{80mm}{!}{\includegraphics[scale=1.2]{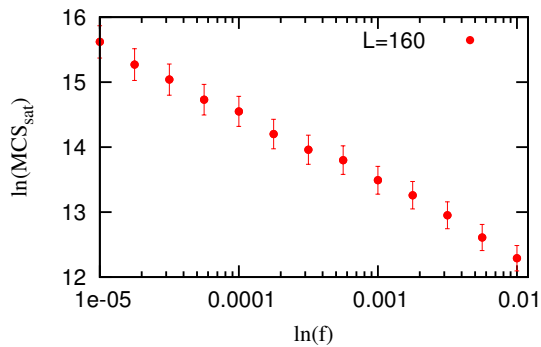}}
\end{center}
\caption{ (Color online) Logarithm of the MC sweeps required to reach the saturation
is plotted against $\ln f$ for $L=160$ system size. The errorbars are shown in the figure.}  
\label{lnfvsmcs}
\end{figure}

\begin{figure}[!h]
\begin{center}
\resizebox{80mm}{!}{\includegraphics[scale=1.2]{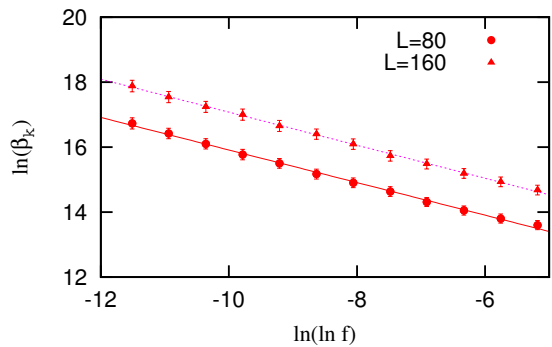}}
\end{center}
\caption{ (Color online) Logarithm of the saturation values of the histogram fluctuations $\ln (\beta_k)$
is plotted against $\ln (\ln f)$ for $L=80$ and $L=160$ systems with the error bars. The
slopes of the two linear fits are given in the text.}
\label{slope}
\end{figure}

In Fig. \ref{hisfluc}, we have plotted the fluctuations in the histogram $\Delta H_k^j$
against the number of MC sweeps $j$ for four values of the modification factor $f$. The
plots shown are for $L=160$ lattice and for $\ln f$ equal to $10^{-2}$, $10^{-3}$, $10^{-4}$ 
and $10^{-5}$. We did not go to values of $\ln f$ less than $10^{-5}$ as it takes a very large
CPU time. Averages were taken over hundred independent simulations to improve the statistics and
accuracy. Similar plots are also taken for the $L=80$ lattice. It is evident from Fig. \ref{hisfluc}
that $\Delta H_k^j$ increases initially and then saturates and as $f$ gets smaller, the 
saturation value as well as the number of MC sweeps necessary to reach the saturation ($\tt MCS_{sat}$)
increases. Fig. \ref{lnfvsmcs} explicitly reveals this fact.
The standard error calculated from the hundred independent simulations are
also shown in Fig. \ref{hisfluc}. In Fig. \ref{slope}, we have plotted the logarithm of saturation value
$\beta_k$, i.e., $\ln(\beta_k)$ vs $\ln (\ln f)$ for system sizes $L=80$ and $L=160$.
From this figure, it is clear that
\begin{equation}
\beta_k \propto (\ln f)^{\alpha}
\label{eqn11}
\end{equation}
where the index $\alpha=-0.50133 \pm 0.007$ for $L=80$ and $\alpha=-0.50844 \pm 0.005$ for
$L=160$ respectively. This is in agreement with the prediction of Zhou and Bhatt \cite{zb}. 
Certainly, this result is not new. It confirms the previous results that the values of the
slope is generic to the WL algorithm, in this case, it is a continuous lattice spin model 
with spin dimensionality three.

\begin{figure}[!h]
\begin{center}
\resizebox{80mm}{!}{\includegraphics[scale=1.2]{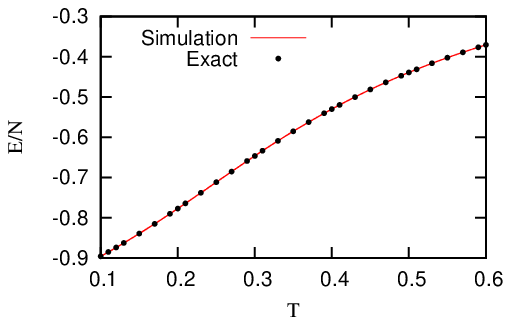}}
\end{center}
\caption{ (Color online) The variation of the average energy per particle is plotted against $T$ (solid line) 
for $L=220$. Exact results, indicated by the filled circle, are also plotted in the same graph.  
The error bars are of the dimensions smaller than the symbols used for plotting.}
\label{engy}
\end{figure}

\begin{figure}[!h]
\begin{center}
\resizebox{80mm}{!}{\includegraphics[scale=1.2]{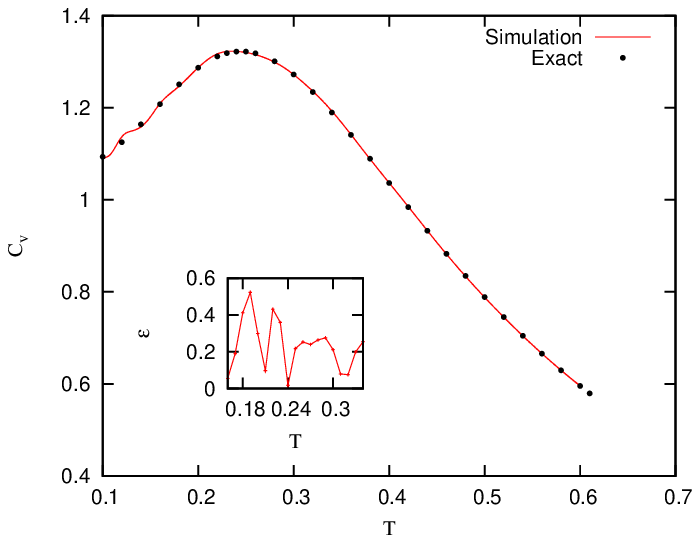}}
\end{center}
\caption{(Color online) The specific heat is plotted against $T$ (solid line) 
for $L=220$. Exact results are indicated by the filled circle.  
The error bars are of the dimensions smaller than the symbols used for plotting.
The percentage error in the $C_v$ in comparison with the exact results is shown in the inset.}
\label{cv}
\end{figure}

\begin{figure}[!h]
\begin{center}
\resizebox{80mm}{!}{\includegraphics[scale=1.2]{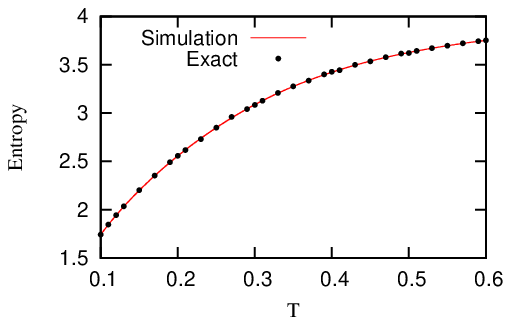}}
\end{center}
\caption{(Color online) The variation of the entropy per particle is plotted against $T$ (solid line) 
for $L=220$. The filled circle indicates the exact results.  
The error bars are of the dimensions smaller than the symbols used for plotting.}
\label{entpy}
\end{figure}

Now we present the results of various thermodynamic quantities obtained from the simulation. 
In Fig. \ref{engy}, we have plotted the average energy per spin against temperature ($T$) for
$L=220$. The results have been compared with the exact values of this observable obtained
from Ref. \cite{vr}.
The specific heat, calculated as fluctuations of the energy, 
has been plotted against $T$ in Fig. \ref{cv} for $L=220$ and compared with
the exact results. In the inset of Fig. \ref{cv}, the percentage error ($\epsilon$) in the $C_v$ 
near the peak in comparison with the exact results is shown.
Percentage error is a measure of how inaccurate (or accurate) a measurement is and is defined by the
formula $\frac{\tt measured~ value - \tt actual~ value}{\tt actual~ value} \times 100 \%$.
Exact results show that the specific heat peak is maximum at a
temperature $T_{max}^{ex}=0.24$ and from our simulation we obtain the temperature at which the peak of the specific heat 
is maximum is $T_{max}^{sim}=0.2351$ for $L=220$. This implies that the percentage error in temperature at 
which the peak of the specific heat is maximum is $2.04 \%$.
Fig. \ref{entpy} shows the variation of entropy per particle for $L=220$ 
and the exact results are also shown in the same plot.

The attention is now focused on the autocorrelation time of the simulation. 
The autocorrelation function for an observable $O(t)$ is given by
\begin{equation}
\chi(t)=\int \mathrm{d}t^{\prime}(O(t^{\prime})-\langle O \rangle)
(O(t^{\prime}+t)-\langle O \rangle)
\label{autocorfn}
\end{equation}
where $O(t)$ is the instantaneous value of the observable at time $t$ and
$\langle O \rangle$ is the average value. The integrand in the above equation
actually measures the correlation between the fluctuation of the
observables at two different times, one an interval $t$ later than
the other. So $\chi(t)$ will take a
nonzero value if on the average the fluctuations are correlated,
otherwise it is zero. Thus when $t$ is just a single MC step
apart, we will have a large positive autocorrelation. For large
$t$, $\chi(t)$ will be zero and the measurements are totally
uncorrelated. The autocorrelation is expected to fall off
exponentially at long times thus:
\begin{equation}
\chi(t) \sim {\tt e}^{-t/\tau}
\label{autocor}
\end{equation}
where $\tau$ is a measure of autocorrelation time of our
simulation. At time $t=\tau$, the autocorrelation function, which
is a measure of the similarity of the two states, is
only a factor of $1/{\tt e}$ down from its maximum value at $t=0$. 
We have estimated the autocorrelation time both for the simulations with the conventional spin
update method and the proposed spin update method. The autocorrelation time is calculated 
following the method proposed by Madras and Sokal \cite{madso}. In the conventional 
spin update scheme, when we flip a single spin in each update, the total energy can only
change by a small amount every time. In the proposed spin update scheme, the change in
total energy is greater compared to that in the conventional scheme. As the WL algorithm does
not require $E_i$ to be close to $E_f$, but compares only $\Omega (E_i)$ and $\Omega (E_f)$, the
convergence becomes faster with the proposed scheme than with the conventional scheme.

\begin{figure}[!h]
\begin{center}
\resizebox{80mm}{!}{\includegraphics[scale=1.2]{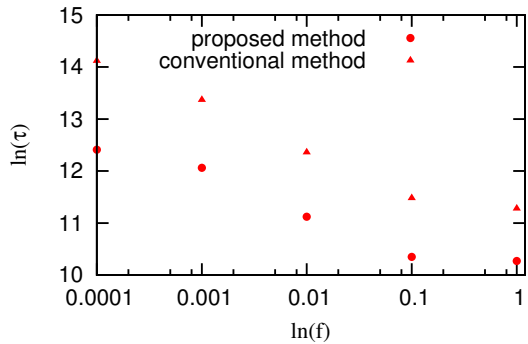}}
\end{center}
\caption{(Color online) Logarithm of the autocorrelation time plotted against $\ln f$ for system size $L=200$
for both the proposed and the conventional spin update method.}  
\label{lnfvstau}
\end{figure}

We have found that the autocorrelation time ($\tau$) exhibits a power law scaling with
system size, i.e., 
\begin{equation}
\tau \propto L^z
\label{eqn12}
\end{equation}
The scaling exponent ($z$) is determined from a linear fit of the plot $\ln \tau$
versus $\ln L$. The logarithm of the autocorrelation time for both the conventional 
and the proposed spin update scheme has been plotted against $\ln L$ for $\ln f=0.01$ 
in Fig. \ref{autocor}. The scaling exponent for the proposed spin update scheme ($z_{\tt new}$) 
is found to be $z_{\tt new}=1.36351 \pm 0.024$ while that for the conventional spin update
scheme ($z_{\tt old}$) is found to be $z_{\tt old}=1.57591 \pm 0.013$. The proposed spin update
scheme significantly decreases the scaling exponent.
\begin{figure}[!h]
\begin{center}
\resizebox{80mm}{!}{\includegraphics[scale=1.2]{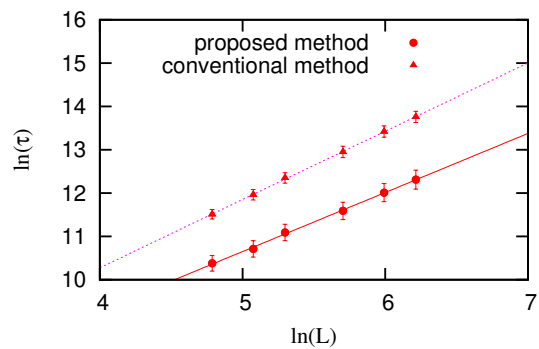}}
\end{center}
\caption{(Color online) Scaling of the autocorrelation time as a function of system size $L$ for 
$\ln f=0.01$ with the error bars shown. The scaling exponents are mentioned in the text.}
\label{autocor}
\end{figure}
We would like to point out that the autocorrelation time ($\tau$) increases rapidly as 
the modification factor ($f$) becomes smaller and for a larger system size, the
calculation of $\tau$, specially for smaller $f$, becomes very much costly in terms
of CPU time. The autocorrelation time for a number of modification factors $f$ for
$L=200$ for both the proposed and the conventional spin update schemes
is listed in Table \ref{table1} and plotted in Fig. \ref{lnfvstau}.
\begin{table}[!h]
\caption{Autocorrelation time (in units of MC sweep) for different $\ln f$ for $L=200$.}
\begin{center}\
\begin{tabular}{|c|c|c|}
\hline
$\ln f$ &$\tau_{\tt new}$ &$\tau_{\tt old}$  \\
\hline
$1.0$ &$28954$ &$78966$ \\
\hline
$0.1$ &$31226$ &$96902$ \\
\hline
$0.01$ &$67642$ &$234502$ \\
\hline
$0.001$ &$173250$ &$638082$ \\
\hline
$0.0001$ &$246118$ &$1359694$ \\
\hline
\end{tabular}
\end{center}
\label{table1}
\end{table}

\section{Conclusions}
\label{conclu}

To summarize, we have tested the performance of the WL algorithm in a continuous lattice spin model, 
namely, the $1d$ LL model which describes a nematic liquid crystal in the mean field approximation. 
The results obtained from our simulation are compared with the exact results available for this 
model. It has been observed that the results obtained tally accurately with the exact results. 
We focus on the fluctuations of histogram and replace the ``flatness'' criterion with
that of minimum histogram. We have found that in this continuous lattice model, the fluctuations in 
the energy histogram, after an initial accumulation stage, saturates to a value that is proportional 
to $1/\sqrt{\ln f}$ where $f$ is the modification factor in the WL algorithm and confirm that this
behavior is generic to the WL algorithm.
We also present a novel method for spin update scheme to obtain a subsequent configuration which is less-correlated than the previous method. 
The proposed spin update scheme makes the WL ``driver'' to move from one sampling point to the next 
faster. As a result, the autocorrelation time between successive moves decreases and
the convergence becomes faster. It may be noted that the WL algorithm only asks for the next 
sampling point (say $X$) with probability distribution $P(X) \propto \Omega(X)/\Omega (\bar X)$ where
$\Omega (X)$ and $\Omega (\bar{X})$ are the exact and the estimated DOS respectively. A previous study
\cite{sbml} suggested that $N$-fold way updates yields better performance in flat-histogram 
sampling. However, Dayal $et. al$ \cite{dtwwtsc} argued that the performance is limited by the added
expense of the CPU time needed to implement the $N$-fold way updates. The proposed method is simple 
to implement and has also the merit that it makes us free from tuning any adjustable
parameter while simulating a continuous lattice spin model. Although the method has been applied
to a liquid crystalline system in the present work, the method can, in general, be applied to any 
lattice spin model with continuous energy spectrum. This method has resulted in efficient 
simulation of continuous models with $XY$ symmetry \cite{sr3,sr4}. Finally, we stress that the focus
in this paper is to test the performance of the WL algorithm in continuous lattice spin models with
the proposed spin update scheme. We hope that this spin update method will be of general interest in 
the area of research in Monte Carlo simulations of continuous lattice spin models.

\section{Acknowledgements}

I wish to thank Prof. S. K. Roy for fruitful discussions and critical reading of the manuscript. 
This work is supported by the UGC Dr. D. S. 
Kothari Post Doctoral Fellowship under grant No. F-2/2006(BSR)/13-398/2011(BSR). Part of the 
computations of this work has been done using the computer facilities of the TCMP Division of 
Saha Institute of Nuclear Physics, Kolkata, India. I thankfully acknowledge the unanimous referee
for a number of suggestions in improving the manuscript.

\end{document}